# Agent-based modeling and the sociology of money: some suggestions for refining monetary theory using social simulation

Eduardo Coltre Ferraciolli* and Tanya V. Araújo

*ISEG Research in Economics and Management, University of Lisbon. Instituto Superiorde Economia e Gestão, Universidade de Lisboa, Lisbon, Portugal.*

* ec.ferraciolli@gmail.com

**Abstract**

The institution of money can be seen as a foundational social mechanism, enabling communities to quantify collectively regulate economic processes – money can be said, indeed, to constitute the micro-macro link in economics. This paper reviews influential views on the nature of money in economics and sociology, contrasting them to the relatively limited findings of recent agent-based models of the "the emergence of money". Noting ample room for novel combinations of sociological and formal methods to drive insight into the many roles played by money in the economy, we conclude by indicating research directions in which we believe this combination can provide new answers to old questions in monetary theory.

**Introduction and Scope**

In this exploratory paper we outline some directions in which we believe social simulation methods could advance theoretical and empirical research on fundamental aspects of the theory of money. Specifically, we will pursue two questions:

- How can agent-based modeling contribute to clarifying the nature of money as an institution, and to what extent should this build upon other formal modeling attempts?
- How can this effort benefit from reflection on the nature of money being conducted mostly within sociologically informed intellectual traditions that are often critical of formal modeling?

The paper is organized as follows: this section lays out scope and objectives. The next section contextualizes the rest of the paper by revisiting classic and recent academic debates on the nature of money: we note positions taken by different intellectual traditions and extract the key concepts that characterize each approach. Section 3 presents a literature review of recent agent-based modeling contributions to this topic: here we outline the general structure of the most important existing models, classify them according to components, strategies and results, and evaluate each from two perspectives: generative epistemology and the "concepts of money" that we identified above. Section 4 then proposes a few promising directions in which we believe social simulation methods might still make substantial contributions to monetary theory. A concluding section reflects on the potential explanatory value, as well as the likely limitations, of deploying agent-based methods to examine foundational questions of economic and institutional organization – many of which are brought into focus through a study of money.

*"The Nature of Money" in 2025*

Investigations into the nature of money have been a relevant - but never primary - concern of both economics and sociology since the disciplines were established at the turn of the 20th century. Interest in this topic, within and across disciplines, has been particularly intense over the past few decades, a period marked by reevaluations of the notion of money in face of developments such as increased financial instability, growing experimentation with alternatives to national currencies, and wide-ranging debates over central bank independence and the role of finance in society. While academic theories about "the nature of money" in economics and sociology form a key part of these discussions, the conceptualizations of money offered by different approaches are markedly different. In fact, caught between macroeconomics' overriding focus on centralized monetary management and microsociology's qualitative investigations of the varied meanings that surround the use of money, conceptual reflection on the actual nature of money - on what kind of institution it is, on how to define it in face of its transformations, on specifying its ontological status as a social form - occupies a secondary and disputed position in both disciplines. This middle ground of fundamental monetary theory - neither microcultural nor macroeconomic, but macrosociological –will be our focus in this article. Over the rest of the paper, we refer to institutions as "integrated systems of rules that structure social interactions" (Hodgson 2015), and we aim to *demarcate* money as a unique and foundational institutional arrangement. We argue that, in combination with social simulation, this view can still generate both valuable insight and useful formal tools for the study of the nature of money and related monetary issues.

At this level of analysis, attempts to answer the question "what is money?" vary widely. This results not only in a variety of conflicting perspectives but also in a series of long-standing, unresolved theoretical disputes. Further complicating matters, a recent part of this debate explicitly attempts to overcome conventional disciplinary divides (Ingham 2004; Ganßmann 2013; Dodd 2018), generating intricate forms of collaboration and contestation across disciplines. With *fiat* money forming a central part of the institutional infrastructure of modern economies, moreover, many of these disputes acquire a discernible political aspect that is often difficult to disentangle from the theories that underlie each position. Far from merely abstract or theoretical debates, then, these divisions are deeply consequential parts of current political economy; their implications play out at different levels, from the grassroots design of local community currencies to the hierarchical organization of currencies in the international monetary system. In fact, divergent conceptualizations of the nature of money can be seen as one of the main fracture lines separating economics and sociology - and creating rifts in the discipline of economics itself – since the field-founding disputes of the early 20th-century; many of the lines drawn then still divide the camps of the debate surrounding monetary theory, and monetary practices, today.

**Echoes of the Methodenstreit: concepts of money between economics and sociology**

Debates surrounding the nature of money can be cast as an ongoing tension dating back at least to the great theoretical debates of the *Methodenstreit*, the confrontation between the Austrian marginalists - with Menger and Mises as central references - and the German Historical School - mostly associated with Knapp and Schmöller, in connection with Simmel, Weber and later Keynes - starting in the 1890s (Frisby 2013). Central to the theme of this paper, one outcome of these disputes was directing the methodology of economics away from historical and sociological observation, and towards more abstraction, formalization, and modeling. These divisions were also consequential in the delimitation of separate disciplinary spaces for economics and sociology, a separation later tacitly validated by Parsons' paradigm (Ingham 2004) and arguably still in effect today.

Money is sometimes considered an "orphan of the *Methodenstreit*", with disciplinary separations arguably leaving both economics and sociology poorly equipped to provide a comprehensive account of its nature as an institution (Ingham 2004; Dodd 2018; Orléan 2020). Traces of the *Methodenstreit* - and irreducible discordance over the nature of money - can also be seen in many of the other major disputes in the history of the social sciences, such as on the socialist calculation debate (Mirowski 2018), in the opposition between metallism and chartalism in monetary theory (Bell 2001), in the formalist-substantivist debate in economic anthropology, on the macroeconomic clashes between monetarism and Keynesianism in the postwar period (Ingham 1996), and around broader divergences regarding the relationship between agency and structure, the causal effect of institutions over individuals, and the nature of social order in general (Orléan 2020; Hodgson 1999). Some version of these divisions is also visible separating the two schools of thought that, as we argue below, are currently producing the most insightful new contributions to the long history of fundamental theories of money: game-theoretical investigations of money as a form of coordination (Bigoni et al. 2020; Araujo & Guimarães 2014; Camera & Gioffré 2014) and the reflections on institutional ontology that take money as their central case (Hodgson 2015; Guala 2020; Herrmann-Pillath 2014). As we suggest over the rest of the paper, one reason for these persistently unresolved theoretical and intellectual difficulties is the fact that money is *constitutive of the micro-macro link in economics*: at once a subjective and systemic phenomenon, one involving both emergence and design, a complex institutional cluster lending itself to many different interpretations.

While it would be impossible to take full stock of these debates here, we outline below some of the most important ways of thinking about the nature of money in different traditions within economics and sociology, identifying (and marking in **bold**) concepts that each approach considers indispensable - and that many other approaches disregard.

*"Models of money" and mainstream economics*

Although much relevant work in mainstream economics attempts to define and explain the fundamentals of money through a descriptive or historical lens (Tobin 1992; Goodhart 1998), the largest and most influential part of this literature approaches the issue through formal models of postulated, abstract economic environments in which money either appears as a solution to specific economic puzzles (Kiyotaki et al. 2016; Camera & Gioffré 2014) or can be said to "play an essential role" (Wallace 2010). These fundamental models can be seen as alternatives to the amply discussed issue of the integration of money in general equilibrium theoretical frameworks, a discussion closely related to the microfoundations project in neoclassical economics (Dopfer et al. 2004). Two older and well-established approaches within these "nature of money" models are traditional "**turnpike** models" with **spatial separation** (Townsend 1980) and **search-theory** models focusing on money as a solution to the **double coincidence of wants problem** in an equilibrium setting (the foundational Kiyotaki & Wright (1993) models, henceforth "KW"). More recent approaches, however, expand their focus to include concerns that overlap with the conventional domain of sociology: mechanism design, for instance, models **transaction costs** that could account for specific aspects of money (Wallace 2010) such as **imperfect monitoring**, **costly connections among people**, and **imperfect recognizability of assets**; an even more recent game theoretical approach (Bigoni et al. 2020; Araujo & Guimaraes 2014; Camera & Gioffré 2014) weaves together themes like **gift-giving** and **reciprocity**, **anonymity** and **social memory**, or **coordination,** into a modeling and experimental perspective. Most, but not all of this literature approaches modeling through the concepts of **rationality** and **equilibrium**. Although arguably sociological in concerns, this recent literature does not appear to reference classic work on the same subject from other social sciences; in any case, it might be considered the richest current incarnation of the winning camp in the post-*Methodenstreit* divides regarding the theory of money.

Many of these approaches refer in one way or another to the early Jevons/Menger descriptions of money as a solution, accomplished through **self-organization**, to the difficulties of barter. A frequent criticism against this position (Ingham 2004; Ganßmann 2013; Graeber 2012) focuses on the a-historicity and implicit market-friendliness of looking at money as a solution to trade frictions. While the variety of views within mainstream economics means that this critique only partially hits its mark, it is still arguably true that the greatest share of formal modeling of "nature of money" issues takes as its starting point a loosely metallist conception of money as primarily a medium of exchange. This raises a central question for the remainder of this paper: to what extent can a formal model – usually restricted to a few specific mechanisms, reliant on possibly arbitrary simplifying assumptions, and primarily logical instead of historical - be considered to *explain* an institution? We suggest that theoretical, exploratory agent-based models of money can answer this challenge by finding grounding both in current institutionalism and in generative epistemology (Epstein2012), in a way that might expand the scope of what fundamental models of money have so far been able to accomplish.

*Old and new references from the social sciences*

The recent work of Geoffrey Ingham (Ingham 2004) has revitalized the macrosociology of money and provided well-documented theoretical and historical support for connections with heterodox economics. Ingham draws on classic sociological thought by Simmel, Weber and Polanyi to present a view of money as an institution that is irreducibly historical, and to draw attention to the **distributive conflict** that always characterizes the production and management of money across societies. The starting point for Ingham's reflection is seeing money as a **"structure of social relations"**, more specifically a set of **anonymized** relations of **credit/debt** or "promises to pay". Ingham's work also underscores the extent to which comprehending money requires an analysis of what we could call its **"institutional embeddedness"**, or the extent to which money can only be understood in relation to an environment composed of other institutions (such as markets, the state, private credit networks, or all of the above in different historical configurations). This perspective presents a sustained challenge to attempts at formal modeling, which tend by necessity to be reductive and focus on one or other closely delimited explanatory mechanisms. Following Keynes' and Schumpeter's positions on the state-and-credit origins of money as well as studies of metrology and of the ethnographic record, Ingham argues that it is the "money of account" and "means of payment" functions of money - thus not "medium of exchange" that are essential to specify the institution. Ingham's work, in its connections with lineages in post-Keynesian economics (Pixley et al. 2013) and political science (Eich 2022), might be considered a current representative of the German Historical School (and of its constitutive interdisciplinarity).

Other recent anthropological studies of money (Maurer 2006; Dodd 2018; Graeber 2012) and finance (Hart & Ortiz 2014), as well as the sociology of complementary currencies (Blanc 2017; Gómez 2018), also draw attention to the extent to which money is an **embedded** part of other economic and non-economic actions rich with meaning, as well as a social practice marked by more **plurality** of manifestations than is usually conceded by economic theory. In contrast, a very different line of thought provides penetrating institutional insight into money's **systemic socioeconomic role** by understanding it, in connection with Parsons and Luhmann, as a **generalized means of communication** (Ganßmann 2013) and as intrinsic to economic modernity: money is here precisely a tool for the **disembeddedness** (Giddens 1990) of human relationships from specific locations or social settings, permitting the **stretching in time and space** of economic activities and enabling increasing **anonymity** in wider and wider systems of mutually coherent economic actions. All of these can be said to follow the spirit and framework laid out by Simmel, as well as Weber's views on rationality and bureaucratization, and express aspects of the institution of money with which economic models do not engage often.

Finally, Simmel and Polanyi, whose contributions go beyond these more recent appropriations, can be considered founders of specifically sociological ways of thinking about money as an institutional form - and can provide theoretical foundations for very different abstract models from the ones offered by economics. Simmel's reflections on money as a paradigmatic form of **objectivity** or **relativity** can be seen as a proposal to understand the economy as a self-organized system - but, in contrast to Menger's, this is a version of self-organization through money in which the role of **representation**, **symbols** and **semiotics** is made explicit. Money is here a very particular shared **idea**, one embodied in **external social forms** and one that enables a form of systemic **economic coordination** that is not necessarily rational or strategic. More than that, money is here **"a claim upon society"** (Simmel 2011), a particular social configuration through which individuals interconnect with each other and with the institutional structures they inhabit. One of Polanyi's most significant contributions, in turn, is a typology of different institutional and organizational forms observed across the anthropological and historical record, built from a perspective that is critical of, but conceptually quite close to, catallaxy. Polanyi's central tools, such as the concepts of **special/general-purpose** money and the "forms of integration" through which societies' disparate parts form **interdependences** that grant coherence to overall economic activity, provide ways to look at social and institutional processes that can't be reduced to other approaches (Polanyi 2011; Codere 1968). Money is at the very core of these bodies of thought - which, as a rule, are not engaged with by most work in economics. More generally, Simmel's notion of social process - as well as Polanyi's related view of the economy as an "instituted process" - provide one possibly privileged entrance point to the "computational modeling of social forms" (Cederman 2005). Money, as a paradigmatic economic institution, is a promising candidate for simulation approaches relying on a sociological perspective.

*Interlocked Economic Heterodoxies*

Post-Keynesian economics and French Monetary Institutionalism, relevant critical traditions in the economic sciences, are closely intertwined with the work in economic sociology, anthropology and political science mentioned above. As mentioned, Geoffrey Ingham's contributions provide a nexus between many of these disciplines, grounding what might be seen a broader interdisciplinary critical alliance between social sciences in the study of money. These traditions emphasize both money's nature as a **credit-debt relation** and the historical role of **coercion** and authority in the historical establishment of money as government-sanctioned tokens for tax payment. Post-Keynesians rely heavily on an understanding of money as a motivator for action and an abstract accounting phenomenon to emphasize the **disruptive potential and instability** intrinsic to the institution, as well as the extent to which money creation happens **endogenously** in the interaction between the productive and financial systems. Definitional concepts to understand the institution of money here are **uncertainty reduction** and socially-sanctioned **liquidity**. The French institutional schools (Orléan 2020), in turn, build their critique of the mainstream upon different sociological and anthropological references (including Mauss and René Girard) to present an account for money's possible **sacrificial origins**, its ability to stand in for society itself as an **external social form**, and the **mimetic**, conventional processes through which the institution is formed; its macroeconomic analysis then privileges issues of the perceived **legitimacy** and potential **fragmentation** of the frail consensus that supports monetary institutions. In different ways, these traditions emphasize the need to understand the specific social and historical formations that sustain the production and management of money, and offer selective recombinations of the works of Keynes, Schumpeter and Marx. A close interplay between these references and work in law and international political economy (Eich 2022; Mehrling 2000; Gilbert & Helleiner 2012) now provides the theoretical foundation for political movements advocating for the "redemocratization of money".

Lying at the other side of the *Methodenstreit*, then and now, are the Austrian and Free Banking Schools in economics. Even more closely connected to the thought of Menger than mainstream economics, these traditions also work closely with linguistics and sociology (Horwitz 1996; Frankel 1977; Selgin 2003), and show great concern with the way institutional or social forms are established - usually seen as an evolutionary, decentralized and largely unconscious process through which fruitful accidents become adopted and reproduce into the future. Money is here, again, a paradigmatic case study for the formation and evolution of institutions. Somewhat paradoxically, the concepts that this tradition tends to emphasize when understanding money are much the same as their post-Keynesian opponents - **trust**, **legitimacy**, **liquidity**, **conventionality**, **instability** and even **reciprocity** (Heering 2005) - but very often in a fully-equipped defense of metallist understandings of money as a commodity used as medium of exchange. Hayek's penetrating insights into the decentralization and self-organization of economic activity inform much of this literature; we would argue that they are both crucial to an understanding of the institution of money and also hampered by this adherence to metallism. We deal with this in more detail in the next section, as many among the most interesting models addressing the fundamentals of money come from authors associated with these traditions.

Perhaps more promising, to conclude, are the variants of heterodox economics that do not subscribe clearly to any of the traditional camps of the post-*Methodenstreit* landscape. Complexity Economics (Arthur 2021) captures much of the Austrian insight on self-organization and institutions, but in a way perhaps less beholden to old theoretical disputes and more explicitly aware of the dangers of reductionism in formal modeling. Some of its key references - including Veblen, Commons amd Simon - are also central to other economic heterodoxies. Finally, institutionalist and evolutionary economics (Hodgson 1999; Dawid 2001; Dosi et al. 2002) combine aspects of all these traditions and deploy tools that are very close to social simulation; we take it as the most promising framework with which to look at money through a rigorous modeling perspective invigorated by sociology. We return to these in the last section.

**Agent-based modeling, economic sociology and "the emergence of money": an overview**

It is precisely at the space of overlap between sociological investigation and abstract, formal modeling of institutions that we want to situate our current reflection on social simulation and the nature of money. If mainstream and Austrian economics provide the field of money studies with sophisticated modeling traditions, the broad heterodox and sociological critiques provide the clearest empirical and historical discussion of money as an institution, and different strands of sociology and anthropology contribute with concepts and perspectives that cannot be reduced to any of the other camps. Agent-based modeling - a naturally interdisciplinary undertaking, much like the theory of money - has a potentially significant contribution to make at the hinge points between these lines of thought. In contrast to modeling approaches relying on general equilibrium and game theory, ABMs are particularly well-suited to handle aspects of institutions that relate to structure, dynamics, emergence and interaction (the latter aligning particularly well with macrosociology's focus on money as a social relation). Theoretical, abstract or exploratory models (as are most of the references we review below) can leverage these features to bring conceptual exploration to the very institutional mechanisms that sociologists emphasize. Empirical, predictive or data-driven models, in turn, relying on the conceptual apparatus and the wealth of qualitative (and increasingly quantitative) data provided by sociological investigations, might find in ABM an appropriate methodological apparatus to study topics rarely addressed by economics. As a relatively recent field marked by both fast change and constant self-reflection, finally, agent-based modeling can also offer epistemological rigor and important counterpoints to older modeling traditions - and a space for renewed reflection on the meaning, potential, and limitations of modeling institutions in general.

However, despite the wide possible scope of agent-based approaches to fundamental questions in the theory of money, most of the existing models so far have addressed what is perhaps the least interesting aspect of the institution: the fact that, among several commodities, one of them may eventually become socially sanctioned as money. In other words, most existing models are illustrations or explorations of Menger's description of the emergence of commodity-money conventions in terms of increasing saleability. We consider this a missed opportunity for social simulation.

*Modeling institutions and "the emergence of money"*

We identify four main approaches to the use of formal agent-based modeling in an attempt to express fundamental concepts regarding the "nature of money", each connecting in varied ways to the traditions that we noted above. Nearly all these are conceptual, illustrative models (Edmonds et al. 2019) with the apparent purpose of supporting a theoretical explanation; although some interact with experimental economics, very few are translatable into forms that can be empirically validated through institutional observation. Most of them make very limited effort to justify assumptions and modeling decisions, providing few foundations on a well-grounded conceptual examination of monetary theories, in a discussion of how abstract models may count as explanations of institutions, or through explicit mappings into the real world. At the end of the section, we present a classification of these models in terms of the key concepts we have flagged above and examine some of their results through the lens of generative epistemology (Epstein 2012).

The first of these model families is broadly in line with mainstream theories of money in economics and draws on the influential three-good, three-agent general-equilibrium KW framework (Kiyotaki & Wright 1993). Several publications have been made following this modelling line (Moran et al. 2013; Rodovalho et al. 2010; Staudinger 1998; Başçı 1999; Kawagoe 2007; Giansante 2006; Hasker & Tahmilci 2008; Yang et al. 2008; Kim 2022; Babutsidze & Iacopetta 2021; Iacopetta 2019), most of them exploring the effect of localized changes in setup or parameters on model results. Foundational to this section of the literature is Marimon et al. (1990)'s application of Holland Classifier Systems to agent-based strategy learning in the KW environment, a contribution that dispenses with some of the assumptions behind equilibrium and rationality modeling, and focuses instead on the *emergence* of a commodity as money – as will many of the following models. This is, however, only a partial break, as the main focus of the simulations is on evaluating whether or not these emergent results converge to the equilibrium conditions defined in KW (for a critique, see Hodgson (2007)). One direction towards which these models have evolved is in combining ABMs, the KW framework and laboratory experiments with human subjects (Duffy 2001; Nioche et al. 2019); these results are relevant but difficult to integrate into a broader understanding of money as a complex institution. Without a more rigorous discussion of modeling purposes, the meaning of emergence, the different non-commodity aspects of money, and of institutions and social processes - all frequently absent from these works - it is difficult to assess whether the contributions truly illuminate core institutional aspects constituting money.

All the models in this tradition, finally, take "media of exchange" as equivalent to money more generally, and can be said to be attached to a commodity-based view of money as a solution to barter frictions; perhaps unsurprisingly, *fiat* money sits in somewhat uneasy coexistence with the rest of the framework (Ganßmann 2013). Despite limitations, in any case, KW and Marimon remain foundational papers in their sections of the modeling literature and can be said to be central references structuring mainstream understandings of money.

A second set of approaches, one less reliant on the KW framework, occasionally critical of it, and showing more internal variety in model setup and findings, draws on Complexity and Austrian Economics and on their common criticisms of the mainstream. Models in these traditions are usually concerned with the insufficiency of some of the neoclassical assumptions that are an intrinsic part of many "fundamental models of money", including the ones in the KW tradition. Agent-based modeling is a natural methodological fit for these perspectives, providing a fresh starting point for conceptual exploration of money that is sensitive to the formation of institutions and of economic structure, to the potential for systemic instability that is implied in social or economic arrangements, and to the way in which it might be properties of agent interaction themselves – instead of any intrinsic feature – that are primarily responsible for the "emergence of money". Gintis (2010) (as well as related models (Gangotena2017)) sets up an environment composed of independent producers-traders and possible product demands (ensuring a lack of double coincidence of wants), then explores how rules of trade between agents may give rise, through self-organization, to something that resembles monetary trade. Relatedly, in a simpler approach, Klein & Selgin (2002) propose an agent-based demonstration of a specific mechanism through which conventions might emerge as self-reinforcing choice process. Although the main theoretical focus of these models is to create artificial representations of economies that operate dynamically far from equilibrium, many of these results can be seen as a generalization of the KW framework allowing wider exploration of parameters and dynamics, and providing a general rationale for the evolution of institutions; while this broadens the discussion considerably, mapping with real-world institutions still remains little discussed and engagement with actual monetary theory is limited. A different, major contribution to this literature is Howitt & Clower (2000)'s account of the twin emergence of both money and productive organization in a decentralized economic environment characterized by trading intermediaries. This is arguably the most well-developed alternative to the KW framework: it builds on Clower's and Leijohnfuvuhd's works on complexity economics and monetary theory, presents a conceptual entry point into the relationship between institutions and economic structure, and addresses some correspondences between formal modeling artifacts and the real economy. Overall, however, this section of the literature offers innovative methods and far-reaching intuitions but does not truly address what it means to model an institution from an agent-based perspective – let alone one as foundational as money. Although focused on emergence as a principle of explanation, little effort is given to specify the concept and the relationship between emergence in abstract models and in real institutions. We view it as an open challenge to social simulation to supersede these important contributions - in particular the implicit metallism that underlies them.

A third family of models relates to theoretical-leaning research published in connection with physics and mathematics journals, especially in Japanese academia, and not often referenced by other works. The most relevant of these is Yasutomi (2003), a discussion on monetary convergence in terms of threshold dynamics with the important result of possible destabilization of previously formed commodity-money conventions. Other meaningful contributions (Kunigami et al. 2010) approach the issue through a "doubly-structural network" approach, where both the commodities space and agent's representations of it evolve. While providing insight and deploying simulation methods not covered by the work mentioned above, these papers tend to be more exploratory and less grounded on the discussions and concepts relevant to monetary theory in either economics or sociology. Connected to this literature, however, important work in Econophysics addresses different aspects or mechanisms of the institution of money through agent-based models, be it general boundary conditions and systemic behavior (Yakovenko 2010) or the instability of money's value (Bak et al. 1999). While not directly related to an exploration of institutional nature, these models have a bearing on it and testify to the flexibility of agent-based methods.

Finally, we also identify occasional attempts to deploy an institutionalist or sociological sensitivity to "nature of money" issues through agent-based modeling; although these are scattered references with limited impact, they come closer to the interdisciplinary point of view we have been defending. Yamadera (2008) extends an analogy of conventional strategy choices to an introductory-level spatial analysis of emergent monetary spaces. Shinohara & Pegio Gunji (2001), in an engaging contribution to learning theory disconnected from other traditions of thought on money, proposes an understanding of monetary conventions emerging from reciprocity providing a possible connection with the macrosociology of money. More than the contributions in themselves, these are relevant as pointers to the possibility of looking at money through a social simulation lens but outside of the commodity-money framework.

*Money, models, coordination, plurality*

As a working reference to drive future reflection, Table 1 presents a summary review of some of these models, evaluating each according to the concepts - among the many possible ones mobilized by different approaches that aim to clarify the nature of money - that they rely on. As the table suggests, there is still ample opportunity for social simulation to provide contributions to the theory of money, especially when informed by an expanded theoretical framework that combines concepts and insights from economics and sociology.

To frame these investigations, it should be said of most of these models that, in the inevitable simplifying assumptions required for formal analysis, there is a risk of losing sight of the most relevant or mysterious features of money as an institution, as well as of key observations from monetary history. Approaching money as an increasingly commonly used commodity has obvious appeal - perhaps too much of it, as this provides a natural, intuitive way to construct models and a possibly misleading mapping into institutions of the real economy. At least for illustrative models, it is much more difficult, and potentially also much more meaningful, to represent money as an institution - as a system of shared rules structuring specific human interactions - and especially as an institution with varying, historically-specific forms. Many of the models also implicitly or explicitly present a process of formal emergence as somewhat equivalent to the historical origin of money, which in turn is presented as evidence towards the nature of the institution. In nearly all these models, then, the underlying epistemology is generative: the fact that something resembling the institution emerges in the model is considered an explanation of that institution. We believe that this is the right framing to ground a rich, renewed understanding of money - as long as it is explicit about the ontology of institutions and not built at the expense of empirical observation of actual monetary forms. What is needed here is to strike a different balance between historical observation and formal models, and we believe that looking at money using the concepts of economic sociology and the tools of social simulation is a promising starting point. The next section explores ways in which this might be accomplished.

Agent-based frameworks and the sociology of money: an invitation, two provocations and some guidelines

In view of the above, we would like to invite closer cooperation between researchers working on both sides of the post-Methodenstreit divides. We outline below five directions in which we believe a combination of agent-based models and historical-sociological thinking can still make distinctive contributions to important open questions in monetary theory. These suggestions are meant to be in line with the epistemology of generative social science (Epstein2012) and the aim of building useful "computational models of social forms" (Cederman 2005). We also indicate, but do not pursue, some of the literature that these investigations could rely on – our intent here is simply to frame important open questions and points of debate on the theory of money in terms that lend themselves to answers deploying social simulation methods.

It should be noted that this might require some of the more insular intellectual traditions involved in fundamental monetary theory to take up the challenge of good-faith engagement. Modelers cannot shy away from thick sociopolitical description of actual institutional history, retreating into the models that are simply the easiest ones to formalize - this remains the case even though some of the more qualitative social studies of money risk taking theory into the exploration of unique instances of money's meanings and non-generalizable reflections on subjectivity. Taking up this challenge also means resisting the simplifying thrust of much abstract theory and conceiving of money in a strong, non-homogenizing sense; an institutionally meaningful account of the nature of money must have something to say not only about coordination over a medium of exchange but also about community currencies, cryptocurrencies, monetary instability and fragmentation, and the great variety of monetary forms in the historical record. Perhaps even more importantly, it must meet minimum criteria of empirical and historical validity - most current models, in either the equilibrium or ABM traditions, fail this test. In turn, researchers aligned with the more qualitative social sciences should not simply dismiss models as useless abstractions - when pressed, they must be able to respond upfront to the challenge of formalizing the mechanisms they consider relevant. This also remains the case even though many models fit almost perfectly the accusation of reductionism, and even though the entire set of extant models, taken as a whole, can be seen as inconsistent in its variety. There are clear social and historical patterns surrounding the emergence, establishment and spread of monetary institutions, and these are at least in principle amenable to formal modeling and to some degree of generalization. Ceding this ground to the most abstract areas of mainstream economics allows modelers to pursue their own, often limited and non-rigorous, uses of empirical evidence - and reinforces old intellectual divisions.

The first four subsections below outline suggestions for the development of primarily exploratory models aiming at illustrating or formalizing insights from sociological theory. A fifth subsection points towards ways in which exploratory models might connect with empirical work. We emphasize theoretical exploration not only because abstract-model discussion is central to how reflection on "the nature of money" takes place in economics, but also because we believe there are particularly valuable contributions from social simulation to be made at this conceptual, foundational level of research - which could later provide new research questions or form the basis for further empirical work. In connection with this, we place special focus on ways in which (inverse) generative social science might provide new ways to address money's role in constituting the micro-macro link in economics, as this carries implications that can extend the scope of fundamental monetary theory.

"The Simmel Challenge" - convergence to common quantification:

While most existing agent-based attempts to illuminate money focus on the medium-of-exchange function and explore a convergence process starting out from a preset space of given commodities, a more relevant research question from a sociologically-informed perspective might be: how it is that different members of a productive community can reach an agreed-upon standard of quantification of value represented through the institution of money? Many of the views we examined above rely on this "nominalist ontology" (Sgambati 2015), a perspective expressed in concepts such as "the working fiction of a monetary invariant" (Mirowski 1991), Keynes' "money-of-account" (Ingham 2004), or "the value of things without the things themselves" (Simmel 2011). Under this perspective, quantification - as a cognitive and social phenomenon - becomes the central question for a theory of money. Of note, this understanding of money as primarily an abstract, institutionalized unit of account is also applicable to social formations where markets do not play a prominent role: as exemplified by Sumerian temple-palace complexes (Hudson 2020) or soviet-style planning (Polanyi 2011), a standardized unit of equivalence of value is also a core element for the centralized allocation of economic resources in redistributive economies. Going beyond commodities and intrinsic features as explanatory principles provides a view of the origin and nature of money that challenges the Mengerian perspective on self-organization through trade, but in a way that also involves convergence and consensus processes for which social simulation could provide relevant analytical and theoretical tools.

One way in which formal models could contribute to this line of thinking about abstract, representational money would be by revisiting the social influence literature (Flache et al. 2017) to investigate collective quantification, exploring mechanisms that could account for the centralized or decentralized establishment of a certain society-wide standard of value, perhaps through a bootstrapping process (Iwai 2011). What in Simmel is a process of objective representation of abstract quantities might be translated, within a model environment, as agent variables representing price estimates for different commodities - and, conversely, for money itself as the base unit of account. Relatedly, once a common standard of value has been established, models in this line might be used to illuminate processes that account for the stability or disruption of this potentially fragile community-validated consensus. Existing agent-based models (Bak et al. 1999; Doepke & Schneider 2017) addressing valuation could be further developed in this direction, and other related theoretical frameworks in the social sciences (Mirowski 1991; Codere 1968; Espeland & Stevens 2003) might be amenable to formalization through ABMs. A second possible modeling contribution could lie in establishing parameter ranges and boundary conditions within which agents' estimates of the value of money and commodities are compatible with social reproduction for a given theoretical production setup - an extension of econophysics' models of money (Yakovenko 2010) more directed to relative price compatibility than to wealth distributions. This second approach also reverts the more common understanding of emergence in an abstract model: modelers would be looking for possible agent-level routines that are compatible with observed aggregate patterns, bringing this effort close to inverse generative social science.

All of this would reconnect monetary theory to broader reflections on value and pricing. Possible empirical applications, if these reflections lead to a conceptual framework that can be precisely captured through a model, could include examinations of inflation and hyperinflation as systemic processes, as well as of the possibilities and limits of centralized, chartal management of the economic unit of account.

"The Polanyi Challenge" - interdependence and economic structure:

Based on ample observation of the historical and anthropological record, Polanyi's work posed important questions and provided innovative conceptual tools (Gregory 2009) with which to understand different socioeconomic formations. General-purpose money, here, is a central component of one specific "form of integration": in contrast to societies whose economy is integrated through redistribution or reciprocity, it is through exchange - thus money and prices - that market economies establish the set of interdependencies between separate economic processes that provide for their coherence and reproduction. These ideas, and the systemic view of the economy that underlies them, are echoed in Giddens' view of money's role in the institutional "disembedding" that removes human activities from specific places and contexts and allows for increased division of interconnected labor spread across time and space. The abstract theme of the formation of economic structure has been explored in the context of social simulation, with (Howitt & Clower 2000) or without (Dawid 2001) a focus on money, but generally only within the narrower remit of agents and commodities. Polanyi's empirically grounded toolbox of contrasted economic and social systems points to a more general study of abstract forms of organization across (Polanyi 2011) or within (Beckert 2009; Granovetter 1985) societies. Under this expanded institutional scope, exploring fundamental monetary theory through structure-formation in agent-based artificial economies potentially becomes part of a more general study of different possible forms of economic organization.

One place to start developing this line of investigation might be in combinations of the Sugarscape tradition with classic models of money focused on production activities that are separated in time and space but must be consistent with each other (Townsend 1980; Howitt & Clower 2000; Marimon et al. 1990). The goal would be to explore in closer detail the conditions under which monetary mechanisms might allow for different levels of specialization and integration of independent economic activities. There are also potentially fruitful interactions here with the toolkit of generative social science (Epstein 2023). A generative framework might take as starting point flexible primitives and concatenations representing economic actions meaningful to different strands of fundamental monetary theory (including, beyond commodities, at least credit-debt and hierarchical agent relations); the theoretical task would then be to explore, for certain definitions of agent welfare and different product-constraining properties of the environment, under which conditions the model gives rise to patterns resembling coherent, self-sustaining social/organizational forms that can reproduce in space and time. Of particular interest would be whether any agent action/interaction patterns emerge that resemble money-mediated exchange, but alternative forms of agent interaction beyond market, price-based trade could form a more general target case: at the very least, the model should be agnostic enough to potentially give rise to patterns of interaction akin to gift-giving and centralized redistribution.

A conceptual framework in these lines could provide relevant theoretical perspective on the viability of different economic systems and forms of economic organization, as well as shed light on the factors that account for the expansion and stability of internally-integrated monetary spaces (including in the study of "the scaling properties of human economies"). This might then be contrasted with historical and anthropological evidence on the evolution of economic organization, including but not limited to the material provided by Polanyi and other works in the tradition of the German Historical School.

Money, intersubjectivity and social norms:

Most of the models mentioned above start off from some notion of individual (often bounded, sometimes perfect) rationality; the role of money then appears as one potential behavior choice after maximization, in (some sort of) equilibrium. This is an aspect of the original search-theory and game-theoretical models that comes under wide criticism, with different contending paradigms emphasizing notions of interdependence (Polanyi 2011), entanglement (Gintis 2017), agent-to-agent as opposed to agent-to-object relations (Ingham 1996), intersubjectivity (Orléan 2020), or some further dimension of social life that cannot be reduced to individuals (Simmel 2011; Herrmann-Pillath 2014; Bell 2001). Simmel's formulation of "money as a claim upon society" captures aspects of this broader view - but, on the whole, there is no straightforward formal equivalent to these ideas that would allow for more comprehensive models of money taking full advantage of these sociological critiques.

One way to overcome these perceived limitations could be to bring fundamental monetary theory closer to current research on social norms (Andrighetto & Vriens 2022). While some connection between these fields has been made in the search-theoretical tradition (Araujo 2004; Araujo & Guimaraes 2014), this has arguably not taken the potential cross-fertilization as far as it can go, as the focus of these models remains on asocial agents and on decisions over commodities instead of in collective action. Other alternatives could involve revisiting well-established theories and models of money but either approaching the issue through non-dyadic agent interaction (Gintis 2017), taking intrinsically social individuals as the starting point (Epstein 2012), or else directly postulating aspects of the model that are external to individual agents (e.g. through global variables that capture something of the idea that the use of money is better understood as a triangular relation between different agents and society). The theoretical interest here would be in what forms of postulated social interaction might account for the dynamics of money as a phenomenon that implicates all individuals as they relate to each other, but that cannot be reduced to any of them - and whether considering this makes a difference to the plausibility of the explanations being proposed.

Even further away from methodological individualism, agent-based models of money might draw on and attempt to formalize well-established holistic conceptions of social order (Gräbner 2016; Orléan 2020). This strong understanding of money as an extrinsic social form - one preceding individuals and which they learn to adopt through inheritance and habituation - could allow modelers to take the institution itself as a unit of analysis; indeed, as the very outcome of the model. This opens up possibilities that, we believe, agent-based models have only begun to explore - particularly in connection with (inverse) generative social science. For instance, Polanyi's notion of "instituted processes" has a clear possible equivalent as action rules shared across model agents (Hodgson 1999; Epstein 2023) reacting to clearly and plausibly defined economic environments. Generative modeling geared towards explicit, meso-level representations of institutions as emergent rule ensembles answers the promise raised by computational social process theory (Cederman 2005) and could provide new perspectives on the classic problem of microfoundations in macroeconomics (Dopfer et al. 2004) - indeed, in a parallel inversion, it might offer ways to explore the macrofoundations of microeconomics.

Closely related would be evaluations of the embeddedness of monetary practices within other social practices and institutions, especially regarding the contrast between general-purpose and special-purpose money (Polanyi 2011), the anthropological view locating money's origins in social practices such as wergeld or sacrifice (Graeber 2012), or the observation of monetary plurality (Blanc 2017). Research along these lines would also address the nature of money through its origins, but in a way that is at least potentially compatible with the anthropological and historical record. Models of this sort, even if purely abstract and conceptual, could also provide perspective on both the rules-in-equilibrium (Guala 2020) and social ontological (Lawson 2018) theoretical paradigms - both of which take money as their central case for the study of institutions.

Money as a form of coordination - or organization:

It is possible to bring together many of the themes discussed above in a more general understanding of the institution of money as a form of coordination. For this, it is useful to distinguish between three different ways to approach the issue. The first one concerns the question of coordination over a medium of exchange: this is the framing adopted by most of models reviewed above, which can be recast in the more abstract form of a simple coordination game wherein it pays off for agents to converge on a shared object as conventional money (Aydinonat 2011). A second set of approaches, the recent focus of theoretical and experimental economics in the "money is memory" tradition (Camera & Gioffré 2014), investigates cooperation through the use of money: adopting a strong interpretation of the fact that digital or paper money possesses no intrinsic value, and viewing the economy as "large interlocking networks of gifts" (Kocherlakota 1998), this line of research finds that the use of intrinsically useless tokens can foster higher levels of cooperation than would be possible without it. Behavioral experiments provide support for some of these propositions (Bigoni et al. 2020).

A third, much broader issue - one echoing Hayek's view of markets as information processors - refers to the fact that monetary mechanisms enable differentiated, independent economic activities, diffused across the economy, to achieve enough internal systemic coherence to become stable and successfully reproduce in time and space: what can be better described as money's role in the coordination of economic activity. This is analogous to seeing collaboration - rather than coordination or cooperation - as a central unexplored object of study organizational theory (Smaldino 2014); it also underlies key work in complexity economics (Clower & Leijonhufvud 1975; Howitt & Clower 2000) and in some of the more interdisciplinary economic heterodoxies (Cartelier 2018; Wilkins & Dragos 2022). As such, a fundamental theoretical understanding of the institution of money would require specifying its role in fostering structured, systemic economic activity - a change of perspective that brings monetary theory back into the realm of major open questions in the history of economic and social thought. This approach to monetary theory would tie together many of the ideas discussed above, including Simmel's focus on abstraction and quantification, Polanyi's emphasis on systemic economic organization, and the explicit modeling of intrinsically social or institutional mechanisms in the regulation of economic activity.

One of the most well-finished expressions of this more ambitious conception of monetary institutions might be Schumpeter's notion of money as "a technology of social accounting" (Peneder 2022): an abstract, society-wide clearing mechanism through which the many individual claims that make up economic transactions are settled and made consistent with each other. This concept finds new purchase in face of digital currencies and blockchain ledgers (Wilkins & Dragos 2022) and resonates with conceptions of agent-based modeling as an exploration of social computing (Epstein 1999). At its most abstract, this is what theoretical research on the "nature of money" has to offer. Agent-based models, as perhaps no other method, are in a position to provide new answers to these questions. Forthcoming work by the authors proposes one way in which this could be accomplished.

Empirical mappings:

As a potential complement for these theoretical exploration models, there is now a host of new sources of historical and contemporary data on monetary practices. Akinobu Kuroda's recent work (Kuroda 2021) redraws the coordinates of the theoretical debates reviewed here and proposes a new framework to identify how different types of social and productive formations might give rise to predictably different forms of money - for instance, how it is that small and proximate productive spaces tend to adopt abstract tokens of small-denomination credit/debt as money, while long-distance trade tends to be associated with anonymous monetary transactions supported by precious metals. Kuroda also presents detailed empirical data on the frequency, type, and scale of transactions for different productive formations in the historical record. Forms of money are here understood in terms of what could be called their "productive embeddedness", forming patterns that can be analyzed, and perhaps generalized, through formal modeling. In connection with the above, the diffusion of specific monetary forms (say, coinage) is also a potentially promising research direction combining agent-based models and historical or current data. The well-established literature on the diffusion of innovation (Kiesling et al. 2012) could serve as a launch point, and new historical databases such as SESHAT (Shin et al. 2020) might provide historical evidence on the emergence and adoption of specific forms of money, informing, for instance, old debates on the role of states and markets in the evolution of monetary practices. Finally, monetary forms complementary to state-and-bank-created modern money are an increasingly well-studied phenomenon. Especially in the case of digital currencies, recent studies relying on network science (Mattsson et al. 2022) have provided a great wealth of high-resolution, individual-level data on money uses and transaction types. In combination with the notion of productive embeddedness, this opens up the possibility that the measurable structure of economic transactions could be predictive of the social form taken by relationships mediated by money. Agent-based models could, at least in principle, be applied here to provide generality and predictive power to these investigations. As somewhat autonomous monetary spaces working at population scales that fit the scope of agent-based methods, complementary currency environments can also provide a testing ground for many of the theoretical concepts discussed above.

**Final remarks: modeling the institutionalization of money**

In the terms of the two research questions raised at the beginning of this paper, we believe the lines of investigation indicated above hold the promise both to address important open questions in monetary theory through new methods and to provide significant use cases that take full advantage of the strengths of agent-based modeling. Previous traditions of formal modeling of money, such as search theory or mechanism design, can offer tools and insight, as well as set out the parameters for the types of questions that might be addressed by rigorous abstract reasoning. Some of the limitations in more traditional modeling can also be superseded by newer methods: social simulation's attention to interaction, process, structure and dynamics, in particular, open up truly new horizons for formal representation and analysis of institutions. No less important, this effort can bring modeling closer to well-established thinking in rich traditions of sociological work that are often disregarded by mainstream economics, providing new ways to operationalize ideas that often only find expression in descriptive or qualitative work. The toolset of social simulation can take monetary theory beyond the traditional remit of model-based economics into a much wider study of general institutional and organizational forms across economic history.

Adopting these methods does not imply logical reductionism, inattention to the historical and anthropological record, or condoning explanations centered on commodities or self-organization - in fact, this type of formal modeling might prove necessary for rigorous exploration of central themes raised by the intellectual traditions that inherit the critical work of the German Historical School. Using these tools well does, however, require a more rigorous relationship with both theory and evidence than has been seen in most recent models of money: modelers must be explicit about whether they are addressing the nature of money or aspects of its operation, whether they are adopting a subjective or a systemic perspective in understanding institutions, whether money is being explained as an object or as a representation, the extent to which the model abstracts away from other embedding social or productive practices, and the specific relationship being claimed between emergence in the model and ideas about the institution's origins. Addressing these would then allow for a more grounded evaluation of the mapping between the model and real-world institutions – preferably one couched in terms of which aspects of money are not covered by the approach employed.

A tension between understanding money as an emergent or a designed institution cuts across many of the intellectual disputes we have considered over this paper. In all likelihood, pure, fundamental monetary theory reproduces rather than resolves these tensions, which arguably can only be overcome within the terms of a broader macrosociological theory of institutions. We believe that models of money can provide valuable contributions to this wider outlook if they explicitly address the micro-macro link, if they are grounded in sociologically-plausible agents, and if they strive for compatibility with both observed institutional history and logical, abstract principles of economic organization. This is a privileged space for the deployment of social simulation. Particularly promising, and tying together many of the points raised in this paper, would be conceptual and empirical models combining inverse generative social science, a sensitivity to the systemically enabling role of money, and formalizations of rule ensembles in line with monetary institutionalism. These would speak directly to current research on experimental game theory and social ontology, but from a naturalistic perspective more in line with the observed history of the institutionalization of money.

| Framework:<br>Concepts of money: | Marimon et al. (mainstream) | Howitt & Clower (complexity) | Gangotena, Klein, Selgin (austrian) | Yasutomi, Kunigami (econophysics) | Yamadera, Orléan (institutionalism) | Promising for social simulation? |
|---|---|---|---|---|---|---|
| *Double coincidence of wants* | yes | yes | yes | yes | no | yes |
| *Transaction costs* | yes | yes | no | no | no | yes |
| *Intrinsic features* | yes? | w/y | no? | no | no | partially |
| *Emergence* | yes | yes | yes | yes | yes | yes |
| *Formation of structure, mediation* | no? | w/y | w/n | no | w/y | yes |
| *Coordination* | partially | w/y | partially | partially | partially | yes |
| *Uncertainty* | no | no? | no | no | w/s | partially |
| *Imitation* | yes | partially | w/n | no | no? | yes |
| *Network effects* | no | no? | w/y | no | no | yes |
| *Path-dependence & Arbitrariness* | yes | partially | yes | yes | yn | yes |
| *Productive specialization* | no | yes | no | no | no | yes |
| *Connectivity* | no | no? | no | no | no | yes |
| *Interdependence, systemic regularities* | partially | partially | w/n | no | no | yes |
| *State token* | no | no | no | no | no | partially? |
| *Anonymity and human economies* | no | no | no | no | no | partially? |
| *Institutionalization rationalization, (dis)Embeddedness* | no | no | no | no | partially | yes? |
| *Memory* | no | no | no | n/w | no | yes |
| *Unit of account, metrology, ideation* | no | no | no | no | no | yes? |
| *Symbols, representation* | no | no | no | nw | no | no |
| *Instability/Disruptive* | no | partially | no | w/n | no | yes |
| *Trust* | no | no | no | no | n/w | yes |
| *Habit, invisibility, legitimacy* | no | no | no | no | w/n | partially? |
| *Fiat, bootstrapping* | no | no | no | no | w/n | yes? |
| *Time-space stretching* | no | no | no | no | no | yes |
| *Plurality* | partially | no | no | no | no | yes |
| *Agent heterogeneity* | partially | partially | w/n | no | no | yes |
| *Material traces, institutionality* | no | no | no | no | no | yes? |
| *Social/distributive conflict* | no | no? | no | no | w/n | partially? |
| *Info asymmetries, recognizability* | no | y/n | n/y | no | no | yes |
| *Institutional embeddedness* | no | no | no | w/n | n/w | yes? |
| *Reciprocity and generalized exchange* | no | no? | no | no | no | yes |
| *Public good, infrastructure* | no | no? | no | no | no | yes |
| *Institutional exteriority, social forms* | partially | n/w | w/n | no | y/n | yes |
| *Meaning, Culture, Violence, Morality* | no | no? | no | no | no | partially? |

Table 1. Evaluation of "nature of money" agent-based models in relation to key concepts in the theory of money across disciplines.

**References:**

Andrighetto, G., and E. Vriens. 2022. "A Research Agenda for the Study of Social Norm Change." *Philosophical Transactions of the Royal Society A: Mathematical, Physical and Engineering Sciences* 380 (2227): 20200411. doi:10.1098/rsta.2020.0411.

Araujo, L. 2004. "Social Norms and Money." *Journal of Monetary Economics* 51 (2): 241–56.

Araujo, L., and B. Guimaraes. 2014. "Coordination in the Use of Money." *Journal of Monetary Economics* 64: 38–46. doi:10.1016/j.jmoneco.2014.01.009.

Arthur, W. B. 2021. "Foundations of Complexity Economics." *Nature Reviews Physics* 3 (2): 136–45. doi:10.1038/s42254-020-00273-3.

Aydinonat, N. E. 2011. "Explaining the Origin of Money: Interdisciplinary Perspectives." In *New Approaches to Monetary Theory: Interdisciplinary Perspectives*.

Babutsidze, Z., and M. Iacopetta. 2021. "The Emergence of Money: Computational Approaches with Fully and Boundedly Rational Agents." *Computational Economics* 58 (1): 3–26.

Bak, P., S. F. Nørrelykke, and M. Shubik. 1999. "Dynamics of Money." *Physical Review E* 60 (3): 2528–32. doi:10.1103/PhysRevE.60.2528.

Başçı, E. 1999. "Learning by Imitation." *Journal of Economic Dynamics and Control* 23 (9-10): 1569–85. doi:10.1016/S0165-1889(98)00084-0.

Beckert, J. 2009. "The Social Order of Markets." *Theory and Society* 38 (3): 245–69. doi:10.1007/s11186-0089082-0.

Bell, S. 2001. "The Role of the State and the Hierarchy of Money." *Cambridge Journal of Economics* 25 (2): 149–63.

Bigoni, M., G. Camera, and M. Casari. 2020. "Money Is More than Memory." *Journal of Monetary Economics* 110: 99–115. doi:10.1016/j.jmoneco.2019.01.002.

Blanc, J. 2017. "Unpacking Monetary Complementarity and Competition: A Conceptual Framework: Table 1." *Cambridge Journal of Economics* 41 (1): 239–57. doi:10.1093/cje/bew024.

Camera, G., and A. Gioffré. 2014. "Game-Theoretic Foundations of Monetary Equilibrium." *Journal of Monetary Economics* 63: 51–63. doi:10.1016/j.jmoneco.2014.01.001.

Carruthers, B. G., and S. Babb. 1996. "The Color of Money and the Nature of Value: Greenbacks and Gold in Postbellum America." *American Journal of Sociology* 101 (6): 1556–91. doi:10.1086/230867.

Cartelier, J. 2018. *Money, Markets and Capital: A Case for Monetary Analysis*. Routledge International Studies in Money and Banking. London ; New York, NY: Routledge, Taylor & Francis Group.

Cederman, L. 2005. "Computational Models of Social Forms: Advancing Generative Process Theory." *American Journal of Sociology* 110 (4): 864–93. doi:10.1086/426412.

Clower, R., and A. Leijonhufvud. 1975. "The Coordination of Economic Activities: A Keynesian Perspective." *The American Economic Review* 65 (2): 182–88.

Codere, H. 1968. "Money-Exchange Systems and a Theory of Money." *Man* 3 (4): 557–77. doi:10.2307/2798579.

Dawid, H. 2001. "On the Emergence of Exchange and Mediation in a Production Economy."

Dodd, N. 2018. "The Social Life of Bitcoin." *Theory, Culture & Society* 35 (3): 35–56. doi:10.1177/0263276417746464.

Doepke, M., and M. Schneider. 2017. "Money as a Unit of Account." *Econometrica* 85 (5): 1537–74. doi:10.3982/ECTA11963.

Dopfer, K., J. Foster, and J. Potts. 2004. "Micro-Meso-Macro." *Journal of Evolutionary Economics* 14 (3): 263–79. doi:10.1007/s00191-004-0193-0.

Dosi, G., L. Marengo, A. Bassanini, and M. Valente. 2002. "Norms as Emergent Properties of Adaptive Learning: The Case of Economic Routines." In *Economic Evolution, Learning, and Complexity*, edited by U. Cantner, H. Hanusch, and S. Klepper, 11–32. Heidelberg: Physica-Verlag HD. doi:10.1007/978-3-642-57646-1_2.

Duffy, J. 2001. "Learning to Speculate: Experiments with Artificial and Real Agents." *Journal of Economic Dynamics and Control* 25 (3-4): 295–319.

Edmonds, B., C. Le Page, M. Bithell, E. Chattoe-Brown, V. Grimm, R. Meyer, C. Montañola-Sales, P. Ormerod, H. Root, and F. Squazzoni. 2019. "Different Modelling Purposes." *Journal of Artificial Societies and Social Simulation* 22 (3): 6.

Eich, S. 2022. *The Currency of Politics: The Political Theory of Money from Aristotle to Keynes*. Princeton: Princeton University Press.

Epstein, J. M. 1999. "Agent-Based Computational Models and Generative Social Science." *Complexity*[1] 4 (5): 41–60. doi:10.1002/(SICI)1099-0526(199905/06)4:5<41::AID-CPLX9>3.0.CO;2-F.[2]

———. 2012. *Generative Social Science: Studies in Agent-Based Computational Modeling:*. Princeton University Press. doi:10.1515/9781400842872.

———. 2023. "Inverse Generative Social Science: Backward to the Future." *Journal of Artificial Societies and Social Simulation* 26 (2): 9.

Espeland, W., and M. Stevens. 2003. "Commensuration as Social Process." *Annual Review of Sociology* 24: 313–43. doi:10.1146/annurev.soc.24.1.313.

Flache, A., M. Mäs, T. Feliciani, E. Chattoe-Brown, G. Deffuant, S. Huet, and J. Lorenz. 2017. "Models of Social Influence: Towards the Next Frontiers." *Journal of Artificial Societies and Social Simulation* 20 (4): 2. doi:10.18564/jasss.3521.

Frankel, S. H. 1977. *Money, Two Philosophies: The Conflict of Trust and Authority*. Oxford: B. Blackwell.

Frisby, D. 2013. *Georg Simmel*. 0 edn. Routledge. doi:10.4324/9780203520185.

Gangotena, S. J. 2017. "Dynamic Coordinating Non-Equilibrium." *The Review of Austrian Economics* 30 (1): 51–82. doi:10.1007/s11138-016-0355-y.

Ganßmann, H. 2013. *Doing Money: Elementary Monetary Theory from a Sociological Standpoint*. First issued in paperback edn. Routledge International Studies in Money and Banking. London New York: Routledge, Taylor & Francis Group.

Giansante, S. 2006. "Social Networks and Medium of Exchange." Working paper. http://andromeda.rutgers.edu/~jmbarr.

Giddens, A. 1990. *The Consequences of Modernity*. Cambridge: Polity Press.

Gilbert, E., and E. Helleiner. 2012. *Nation-States and Money: The Past, Present and Future of National Currencies*. London: Routledge.

Gintis, H. 2010. "The Dynamics of Generalized Market Exchange." Santa Fe Institute Working Paper.

———. 2017. *Individuality and Entanglement: The Moral and Material Bases of Social Life*. Princeton: Princeton University Press.

Goodhart, C. A. E. 1998. "The Two Concepts of Money: Implications for the Analysis of Optimal Currency Areas." *European Journal of Political Economy* 14 (3): 407–32. doi:https://doi.org/10.1016/S0176-2680(98)00015-9.

Graeber, D. 2012. *Debt: The First 5,000 Years*. 10th anniversary edition edn. Brooklyn, NY: Melville House.

Granovetter, M. 1985. "Economic Action and Social Structure: The Problem of Embeddedness." *American Journal of Sociology* 91 (3): 481–510. doi:10.1086/228311.

Gregory, C. 2009. "Whatever Happened to Householding?" In *Market and Society: The Great Transformation Today*. Cambridge University Press.

Gräbner, C. 2016. "Agent-Based Computational Models– a Formal Heuristic for Institutionalist Pattern Modelling?" *Journal of Institutional Economics* 12: 241–61. doi:10.1017/S1744137415000193.

Guala, F. 2020. "Money as an Institution and Money as an Object." *Journal of Social Ontology* 6 (2): 265–79. doi:10.1515/jso-2020-0028.

Gómez, G. M. 2018. "The Monetary System as an Evolutionary Construct." In *Monetary Plurality in Local, Regional and Global Economies*. Routledge.

Hart, K., and H. Ortiz. 2014. "The Anthropology of Money and Finance: Between Ethnography and World History." *Annual Review of Anthropology* 43 (1): 465–82. doi:10.1146/annurev-anthro-102313-025814.

Hasker, K., and A. Tahmilci. 2008. "The Rise of Money: An Evolutionary Analysis of the Origins of Money." Working paper. http://www.bilkent.edu.tr/~hasker/Research/Hasker-Tahmilci-evolution-of-money08-05-15.pdf.

Heering, W. W. 2005. "Money and Reciprocity in the Extended Order.an Essay on the Evolution and Cultural Function of Money." *Entrepreneurship, Money and Coordination*, 156.

Herrmann-Pillath, C. 2014. "Naturalizing Institutions: Evolutionary Principles and Application on the Case of Money." *Jahrbücher für Nationalökonomie und Statistik* 234 (2-3): 388–421. doi:10.1515/jbnst-2014-2-315.

Hodgson, G. M. 1999. *Evolution and Institutions*. Edward Elgar Publishing.

———. 2007. "Institutions and Individuals: Interaction and Evolution." *Organization Studies* 28 (1): 95–116. doi:10.1177/0170840607067832.

———. 2015. "On Defining Institutions: Rules versus Equilibria." *Journal of Institutional Economics* 11 (3): 497–505. doi:10.1017/S1744137415000028.

Horwitz, S. 1996. "Money, Money Prices, and the Socialist Calculation Debate." In *Advances in Austrian Economics*, vol. 3, 59–77. Emerald Group Publishing Limited. doi:10.1016/S15292134(96)03005-0.

Howitt, P., and R. Clower. 2000. "The Emergence of Economic Organization." *Journal of Economic Behavior & Organization* 41 (1): 55–84. doi:10.1016/S0167-2681(99)00087-6.

Hudson, M. 2020. "Origins of Money and Interest: Palatial Credit, Not Barter." In *Handbook of the History of Money and Currency*, edited by S. Battilossi, Y. Cassis, and K. Yago, 45–65. Singapore: Springer. doi:10.1007/978-981-13-0596-2_1.

Iacopetta, M. 2019. "The Emergence of Money: A Dynamic Analysis." *Macroeconomic Dynamics* 23 (07): 2573–96. doi:10.1017/S1365100517000815.

Ingham, G. 1996. "Money Is a Social Relation." *Review of Social Economy* 54 (4): 507–29.

———. 2004. *The Nature of Money*. Cambridge, UK ; Malden, MA: Polity.

Iwai, K. 2011. "The Second End of Laissez-Faire: The Bootstrapping Nature of Money and the Inherent Instability of Capitalism." In *New Approaches to Monetary Theory: Interdisciplinary Perspectives*. Available at SSRN 1861949.

Kawagoe, T. 2007. "Learning to Use a Perishable Good as Money." In *Multi-Agent-Based Simulation VII*, edited by L. Antunes and K. Takadama, 96–111. Lecture Notes in Computer Science. Berlin, Heidelberg: Springer. doi:10.1007/978-3-540-76539-4_8.

Kiesling, E., M. Günther, C. Stummer, and L. M. Wakolbinger. 2012. "Agent-Based Simulation of Innovation Diffusion: A Review." *Central European Journal of Operations Research* 20 (2): 183–230. doi:10.1007/s10100-011-0210-y.

Kim, J. H. 2022. "Emergence of a Good as a Medium of Exchange in Different Types of Networks." doi:10.2139/ssrn.4037872.

Kiyotaki, N., R. Lagos, and R. Wright. 2016. "Introduction to the Symposium Issue on Money and Liquidity." *Journal of Economic Theory* 164: 1–9. doi:10.1016/j.jet.2016.03.012.

Kiyotaki, N., and R. Wright. 1993. "A Search-Theoretic Approach to Monetary Economics." *The American Economic Review*, 63–77.

Klein, P. G., and G. Selgin. 2002. "11 Menger's Theory of Money: Some Experimental Evidence." *What Is Money?* 6: 217.

Kocherlakota, N. R. 1998. "Money Is Memory." *Journal of Economic Theory* 81 (2): 232–51. doi:10.1006/jeth.1997.2357.

Kunigami, M., M. Kobayashi, S. Yamadera, T. Yamada, and T. Terano. 2010. "A Doubly Structural Network Model: Bifurcation Analysis on the Emergence of Money." *Evolutionary and Institutional Economics Review* 7 (1): 65–85. doi:10.14441/eier.7.65.

Kuroda, A. 2021. *Global History of Money*. First issued in paperback 2021 edn. Routledge Explorations in Economic History. London New York: Routledge Taylor & Francis Group.

Lawson, T. 2018. "The Constitution and Nature of Money[A Critique of Lawson's 'Social Positioning and the Nature of Money']." *Cambridge Journal of Economics* 42 (3): 851–73.

Marimon, R., E. McGrattan, and T. J. Sargent. 1990. "Money as a Medium of Exchange in an Economy with Artificially Intelligent Agents." *Special Issue on Computer Science and Economics* 14 (2): 329–73. doi:10.1016/0165-1889(90)90025-C.

Mattsson, C. E. S., T. Criscione, and F. W. Takes. 2022. "Circulation of a Digital Community Currency." doi:10.48550/arXiv.2207.08941. arXiv:2207.08941 [physics, q-fin].

Maurer, B. 2006. "The Anthropology of Money." *Annual Review of Anthropology* 35 (1): 15–36. doi:10.1146/annurev.anthro.35.081705.123127.

Mehrling, P. 2000. "Modern Money: Fiat or Credit?" *Journal of Post Keynesian Economics* 22 (3): 397–406. doi:10.1080/01603477.2000.11490247.

Mirowski, P. 1991. "Postmodernism and the Social Theory of Value." *Journal of Post Keynesian Economics* 13 (4): 565–82.

———. 2018. "Polanyi vs Hayek?" *Globalizations* 15 (7): 894–910. doi:10.1080/14747731.2018.1498174.

Moran, T., M. Brede, A. Ianni, and J. Noble. 2013. "The Origin of Money: An Agent-Based Model." In *Advances in Artificial Life, ECAL 2013*, 472–79. MIT Press. doi:10.7551/978-0-262-31709-2-ch068.

Nioche, A., B. Garcia, G. Lefebvre, T. Boraud, N. P. Rougier, and S. Bourgeois-Gironde. 2019. "Coordination over a Unique Medium of Exchange Under Information Scarcity." *Palgrave Communications* 5 (1): 153. doi:10.1057/s41599-019-0362-2.

Orléan, A. 2020. "Money: Instrument of Exchange or Social Institution of Value?" In *Institutionalist Theories of Money: An Anthology of the French School*, edited by P. Alary, J. Blanc, L. Desmedt, and B. Théret, 239–64. Cham: Springer International Publishing. doi:10.1007/978-3-030-59483-1_8.

Peneder, M. 2022. "Digitization and the Evolution of Money as a Social Technology of Account." *Journal of Evolutionary Economics* 32 (1): 175–203. doi:10.1007/s00191-021-00729-4.

Pixley, J., G. C. Harcourt, and G. K. Ingham, eds. 2013. *Financial Crises and the Nature of Capitalist Money: Mutual Developments from the Work of Geoffrey Ingham*. Basingstoke: Palgrave Macmillan.

Polanyi, K. 2011. "The Economy as Instituted Process." In *The Sociology of Economic Life*. 3 edn. Routledge.

Rodovalho, W. M., C. D. N. Vinhal, and G. D. Cruz. 2010. "Studying the Emergence of Money by Means of Swarm Multi-Agent Simulation." In *Ibero-American Conference on Artificial Intelligence*, 296–305. Springer, Berlin, Heidelberg.

Selgin, G. 2003. "Adaptive Learning and the Transition to Fiat Money." *The Economic Journal* 113 (484): 147–65. doi:10.1111/1468-0297.00094.

Sgambati, S. 2015. "The Significance of Money Beyond Ingham's Sociology of Money." *European Journal of Sociology / Archives Européennes de Sociologie* 56 (2): 307–39. doi:10.1017/S0003975615000144.

Shin, J., M. H. Price, D. H. Wolpert, H. Shimao, B. Tracey, and T. A. Kohler. 2020. "Scale and Information-Processing Thresholds in Holocene Social Evolution." *Nature Communications* 11 (1): 2394. doi:10.1038/s41467-020-16035-9.

Shinohara, S., and Y. Pegio Gunji. 2001. "Emergence and Collapse of Money Through Reciprocity." *Applied Mathematics and Computation* 117 (2-3): 131–50. doi:10.1016/S0096-3003(99)00169-1.

Simmel, G. 2011. *The Philosophy of Money*. Routledge Classics. Abingdon, Oxon ; New York: Routledge.

Smaldino, P. E. 2014. "The Cultural Evolution of Emergent Group-Level Traits." *Behavioral and Brain Sciences* 37 (3): 243–54. doi:10.1017/S0140525X13001544.

Staudinger, S. 1998. "Money as Medium of Exchange - An Analysis with Genetic Algorithms." *IFAC Proceedings Volumes* 31 (16): 93–98. doi:10.1016/S1474-6670(17)40464-2.

Tobin, J. 1992. "Money." Cowles Foundation Discussion Papers.

Townsend, R. M. 1980. "Models of Money with Spatially Separated Agents." In *Models of Monetary Economies*, 265–303. Minneapolis: Federal Reserve Bank of Minneapolis.

Wallace, N. 2010. "Chapter 1 - The Mechanism-Design Approach to Monetary Theory." In *Handbook of Monetary Economics*, edited by B. M. Friedman and M. Woodford, 3:3–23. Elsevier. doi:10.1016/B978-0-44453238-1.00001-6.

Wilkins, I., and B. Dragos. 2022. "Money as a Computational Machine." *Finance and Society* 8 (2): 110–28. doi:10.2218/finsoc.7762.

Yakovenko, V. M. 2010. "Statistical Mechanics Approach to the Probability Distribution of Money." doi:10.48550/ARXIV.1007.5074.

Yamadera, S. 2008. "Examining the Myth of Money with Agent-Based Modelling." In *Social Simulation: Technologies, Advances and New Discoveries*, 252–63. IGI Global.

Yang, J.-S., O. Kwon, W.-S. Jung, and I.-m. Kim. 2008. "Agent-Based Approach for Generation of a Money-Centered Star Network." *Physica A: Statistical Mechanics and Its Applications* 387 (22): 5498–5502. doi:10.1016/j.physa.2008.05.025.

Yasutomi, A. 2003. "Itinerancy of Money." *Chaos: An Interdisciplinary Journal of Nonlinear Science* 13 (3): 1148–64. doi:10.1063/1.1604593.